\documentclass{PoS}

\usepackage{amsmath,amssymb}
\usepackage{graphicx}

\usepackage{setspace}
\usepackage{color}

\title{Free-form Smeared Bottomonium Correlation Functions}

\ShortTitle{Free-form Smeared Bottomonium}

\author{\speaker{Mark Wurtz}\\
        Department of Physics and Astronomy, York University, Toronto, Ontario, Canada M3J 1P3\\
        E-mail: \email{mwurtz@yorku.ca}}

\author{Randy Lewis\\
        Department of Physics and Astronomy, York University, Toronto, Ontario, Canada M3J 1P3\\
        E-mail: \email{randy.lewis@yorku.ca}}

\author{R.~M.~Woloshyn\\
        TRIUMF, 4004 Wesbrook Mall, Vancouver, British Columbia, Canada V6T 2A3\\
        E-mail: \email{rwww@triumf.ca}}

\abstract{Gauge-invariant sources with a hydrogen wave function shape are constructed for bottomonium two-point correlation functions using the free-form smearing technique. The bottomonium spectrum, including a first lattice result for the D-wave first-excited state, is extracted from free-form smeared correlation functions. Results are compared with conventional smearing techniques and free-form smearing is found to have the advantage of reduced statistical errors.}

\FullConference{The 32nd International Symposium on Lattice Field Theory,\\
		23-28 June, 2014\\
		Columbia University New York, NY}

\begin{document}

\section{Introduction}\label{sec:intro}

Free-form smearing is a new gauge-invariant smearing technique that was originally applied to relativistic quarks \cite{vonHippel:2013yfa}. We have applied this method for the first time to the case of non-relativistic heavy quarks to extract the spectrum of bottomonium and compared its performance with the conventional smearing methods of NRQCD: gauge-invariant gaussian smearing and gauge-fixed wavefunction smearing.

\section{Conventional Smearing Methods for NRQCD}\label{sec:conventional}

Gauge-invariant gaussian smearing is a popular quark field smearing method which enhances the ground state contribution to a correlation function and suppresses contributions from excited states. It is an iterative method given by
\begin{align}
\tilde{\psi}_y(x)=\left[1+\frac{\alpha}{n}\Delta\right]^n\psi(y) \quad , \label{gaussiansmear}
\end{align}
where $\Delta$ is the discrete gauge-covariant Laplacian operator, which contains gauge link variables.

Another commonly used smearing technique for heavy non-relativistic quarks involves (Coulomb) gauge-fixing the links and smearing the quark field to an arbritary shape of one's choice \cite{Davies:1994mp}:
\begin{align}
\tilde{\psi}(x)=\sum_y f(x-y)\,\psi(y) \quad . \label{wavesmear}
\end{align}
A function $f(x-y)$ which resembles the wave function of a physical state can enhance the ground state signal. The function $f(x-y)$ can also be chosen to suppress the ground state, allowing for a much cleaner excited state signal. Gauge-fixing is required since Eq.~\eqref{wavesmear} is not gauge-invariant.

\section{Free-form Smearing Method}\label{sec:freeform}

The free-form smearing method \cite{vonHippel:2013yfa} combines the advantages of the conventional methods. It allows one to construct a source with an arbitrary shape while retaining gauge invariance. This is accomplished by iteratively applying the gauge-invariant gaussian smearing method as in Eq.~\eqref{gaussiansmear} to a point source at point $y$ so that gauge link paths connect to every point $x$ on the source time slice. The average of the norm of the source $\left<\left\|\tilde{\psi}_y(x)\right\|\right>=\left<\sqrt{\operatorname{Tr}\left(\tilde{G}^\dag_y(x)\tilde{G}_y(x)\right)}\right>$, where $\tilde{G}_y(x)$ is the gaussian smeared heavy quark propagator at the source time step and the trace is over spin and colour, is used to divide out the approximate gaussian shape of $\tilde{\psi}_y(x)$ and leave a flat distribution with small fluctuations. An arbitrary shape is then applied by simple multiplication of a function $f(x-y)$, and the free-form smearing operation is given by
\begin{align}
\tilde{\tilde{\psi}}_y(x)= \frac{\tilde{\psi}_y(x)}{\left<\left\|\tilde{\psi}_y(x)\right\|\right>} \, f(x-y) \quad .
\end{align}
A significant disadvantage of free-form smearing in its present form is that it is not computationally feasible to apply it at the sink. The reason is that one is required to smear every point $y$ separately and then perform a summation to obtain a momentum projection.

The gaussian smearing parameters used in the free-form smearing procedure were $\alpha=0.15$ and $n=64$. The free-form smeared correlation functions were fairly insensitive to changes in $\alpha$ and $n$, so long as $\frac{\alpha}{n}$ is not too large. Stout smeared links (parameters $\rho=0.15$ and $n_\rho=10$ as defined in \cite{Morningstar:2003gk}) were also used, but this only produced a minor improvement.

For bottomonium, hydrogen-like wave functions have been used with the gauge-fixed smearing method \cite{Gray:2005ur} and were found to be quite effective. For this study, we use functions of the form
\begin{align}
\text{S-wave:}& \quad f(x)=\left\{
\begin{array}{l}
e^{-\frac{r}{a}} \\
(r-b)\, e^{-\frac{r}{a}} \\
(r-c)(r-b)\, e^{-\frac{r}{a}} 
\end{array} \right. \label{swave} \\
\text{P-wave:}& \quad f_i(x)=\left\{
\begin{array}{l}
\tilde{x}_i\, e^{-\frac{r}{a}} \\
\tilde{x}_i\, (r-b)\, e^{-\frac{r}{a}}
\end{array} \right. \label{pwave} \\
\text{D-wave:}& \quad f_{ij}(x)=\left\{
\begin{array}{ll}
\tilde{x}_i \tilde{x}_j\, e^{-\frac{r}{a}} \\
\tilde{x}_i \tilde{x}_j\, (r-b)\, e^{-\frac{r}{a}}
\end{array} \right. \label{dwave} \\
\text{F-wave:}& \quad f_{ijk}(x)=\tilde{x}_i \tilde{x}_j \tilde{x}_k\, e^{-\frac{r}{a}} \\
\text{G-wave:}& \quad f_{ijkl}(x)=\tilde{x}_i \tilde{x}_j \tilde{x}_k \tilde{x}_l\, e^{-\frac{r}{a}}
\end{align}
where $r=\sqrt{x_1^2+x_2^2+x_3^2}$, $\tilde{x}_i = \sin\left(\frac{2\pi x_i}{L}\right)$ and the parameters $(a,b,c)$ are tuned individually to obtain an optimal signal for the ground state, first excited state and, for the S-wave, even the second excited state. A selection of optimized free-form smearing parameters is shown in Table~\ref{table-hydrogen}.

\begin{table}[htb]
\centering
\caption{Examples of optimized free-form smearing parameters from Eqs.~\protect\eqref{swave}, \protect\eqref{pwave} and \protect\eqref{dwave}.}
\label{table-hydrogen}
\begin{tabular}{c|ccc|ccc|c}
\hline
&& ground state &&& first-excited state && second-excited\\
& S-wave & P-wave & D-wave & S-wave & P-wave & D-wave & S-wave\\
\hline
a & 1.6 & 2.0 & 2.5 & 2.8 & 3.0 & 3.5 & 3.0\\
b &     &     &     & 2.8 & 4.5 & 6.5 & 2.13\\
c &     &     &     &     &     &     & 6.0
\end{tabular}
\end{table}

\begin{figure}[htb]
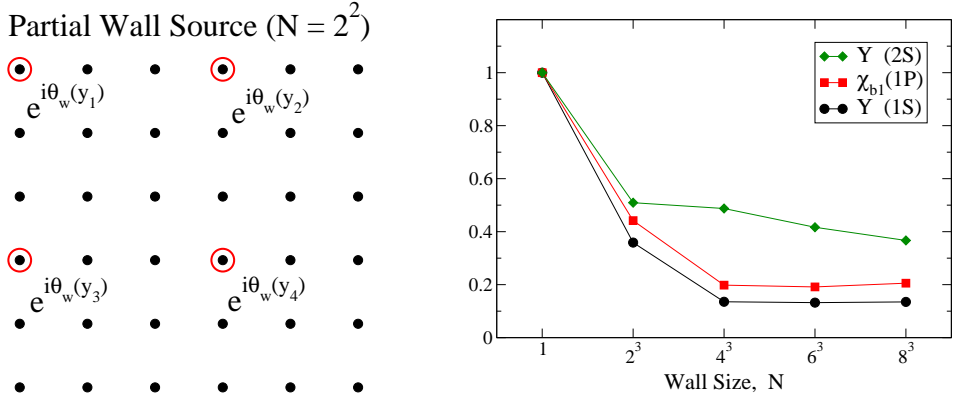

\centering
\includegraphics[scale=0.28]{partwall.eps}
\hspace{12mm}\includegraphics[scale=0.28]{freeform_walltest-norm.eps}
\caption{Left frame: illustration of a random $U(1)$ partial wall source. Right frame: reduction of statistical errors of the $\Upsilon$ S-wave ground state and first-excited state and the $\chi_{b1}$ P-wave ground state energies as a function of the number of points in the free-form smeared partial wall source.}
\label{figure-wall}
\end{figure}

Random $U(1)$ wall sources can often lead to significant reductions in statistical uncertainties of meson correlation functions. The gaussian and gauge-fixed smearing methods given by Eqs.~\eqref{gaussiansmear} and \eqref{wavesmear} can be applied trivially to a full random wall source. For the same reason that it is not feasible to apply free-form smearing to the sink, it is very computationally expensive to implement a free-form smeared full wall source. One would have to smear every point $y$ independently, multiply each by a random unit complex number and sum the results:
\begin{align}
\tilde{\tilde{\psi}}_w(x)=\sum_i^N e^{i\theta_w(y_i)}\, \tilde{\tilde{\psi}}(x;y_i) \quad .
\end{align}
However, to obtain reduced statistical errors of the same quality as a full wall source it is sufficient to use a partial wall source as illustrated in Fig.~\ref{figure-wall}. Significant reductions in statistical uncertainties were found for ``sparse'' $2^3$ and $4^3$ sized wall sources and very little improvement when more points were included in the wall. A $4^3$ sized partial wall source was used to obtain the free-form smeared results presented in the sections below. This wall size was not a computational burden and the improvement was well worth the effort. The gaussian and gauge-fixed smearing methods presented below used a full wall source.

A gauge field ensemble from the PACS-CS Collaboration was used for this study \cite{Aoki:2008sm}: Iwasaki gauge action, clover-Wilson fermion action, 198 configurations, $32^3\times 64$, $a=0.0907(13)\text{fm}$, $n_f=2+1$ and $m_\pi =156(7)\text{GeV}$. The NRQCD action consists of ${\cal O}(v^4)$ terms, tree level coefficients $c_i=1$ ($0\leq i\leq 6$), tadpole improved mean link in Landau gauge $u_L=0.8463$, bare NR-quark mass $M_b=1.95$ and stability parameter $n=4$. The gauge-field ensemble and NRQCD action were chosen to be the same as in \cite{Lewis:2012ir}.

\section{Smearing Method Comparision}\label{sec:compare}

\begin{figure}[htb]
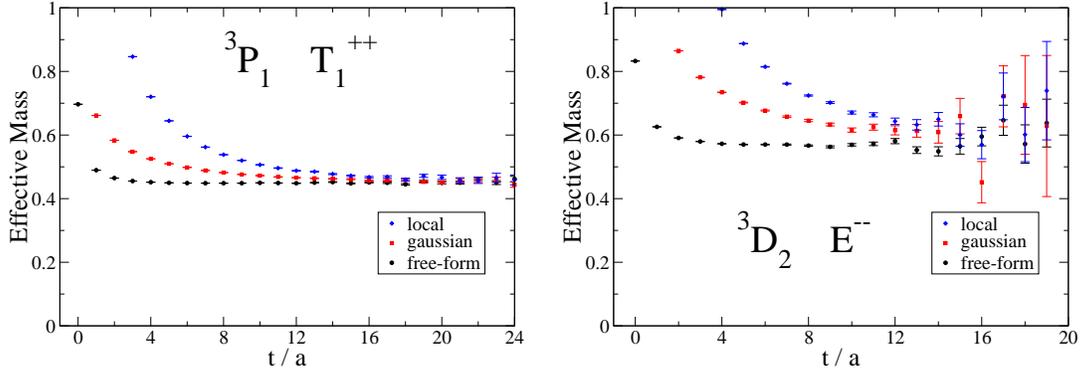

\centering
\includegraphics[scale=0.28]{effmass_gaussian_ground_3P1_T1++.eps}
\hspace{4mm}\includegraphics[scale=0.28]{effmass_gaussian_ground_3D2_E--.eps}
\caption{Effective mass plots of free-form smeared and gauge-invariant gaussian smeared P and D-wave bottomonium correlation functions. A local operator is also shown for comparison.}
\label{figure-gaussian}
\end{figure}

The gauge-invariant gaussian smearing method enhances the ground state signal and suppresses excited states. Figure~\ref{figure-gaussian} shows a comparison of gaussian smearing and free-form smearing optimized for the ground state for select P and D-waves of bottomonium. The effective mass plateaus to the ground state much earlier for free-form smearing than for gaussian smearing. A local operator illustrates the overall improvement. Smearing is applied to the source but not the sink to allow a fair comparison. Also, smearing non-relativistic heavy quarks at the sink increases statistical errors, which was also observed in Figs.~3 and 4 of \cite{Davies:1994mp}.

\begin{figure}[htb]
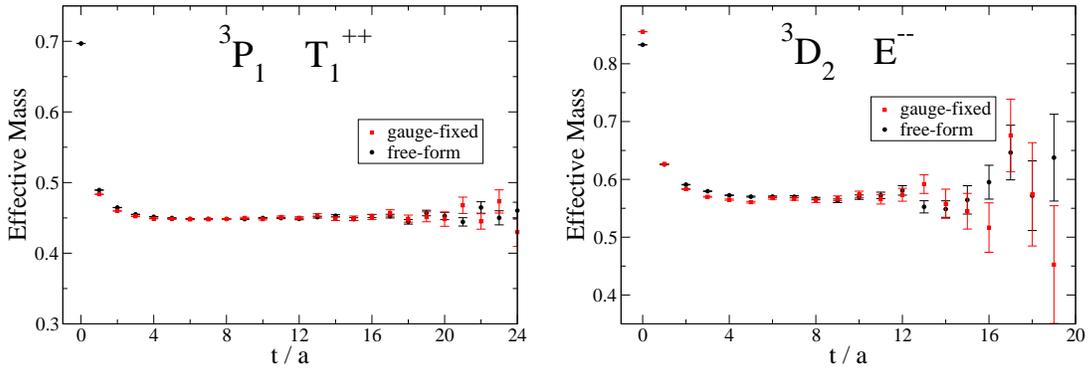

\centering
\vspace{6mm}
\includegraphics[scale=0.28]{effmass_gaugefix_ground_3P1_T1++.eps}
\hspace{4mm}\includegraphics[scale=0.28]{effmass_gaugefix_ground_3D2_E--.eps}
\caption{Effective mass plots for the free-form smeared and gauge-fixed wave function smeared S, P and D-wave bottomonium correlation functions where both methods were tuned to optimize the ground state.}
\label{figure-gaugefix-ground}
\end{figure}

\begin{figure}[htb]
\centering
\includegraphics[scale=0.28]{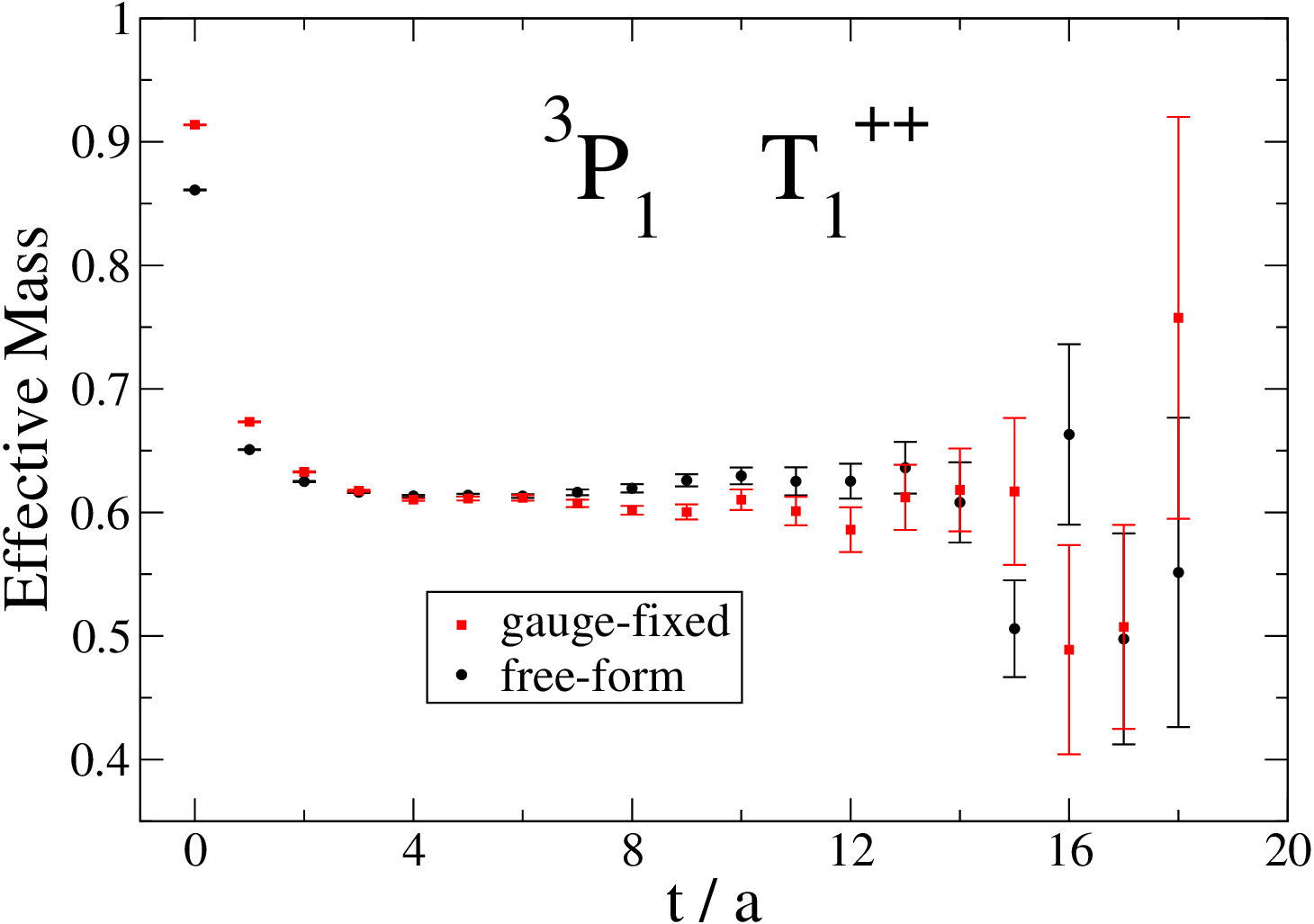}
\hspace{4mm}\includegraphics[scale=0.28]{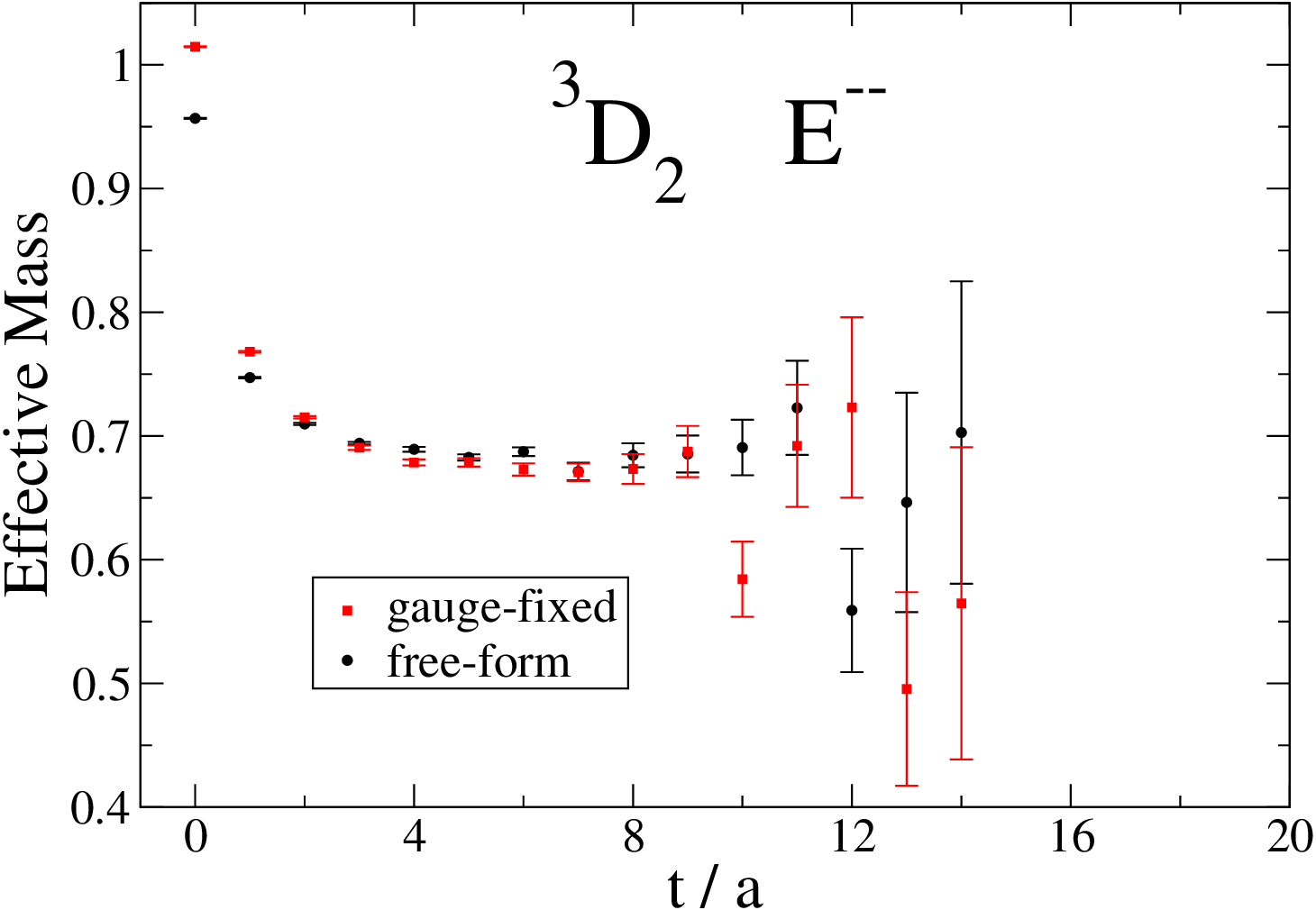}
\caption{Effective mass plots for the free-form smeared and gauge-fixed wave function smeared P and D-wave bottomonium correlation functions where both methods were tuned to optimize the first-excited state.}
\label{figure-gaugefix-excited}
\end{figure}

Figure~\ref{figure-gaugefix-ground} compares effective masses for the gauge-fixed wave function smearing and free-form smearing methods for P and D-wave bottomonium, where both methods were tuned to optimize the ground state. Plateaus occur at roughly the same number of time steps for both methods. Effective mass plots of the two methods for P and D-wave bottomonium optimized for the first-excited state are shown in Fig.~\ref{figure-gaugefix-excited}. The plateaus are at a higher energy than in Fig.~\ref{figure-gaugefix-ground}, indicating that the ground state contribution has been suppressed. Although it is not obviously visible in Figs.~\ref{figure-gaugefix-ground} and \ref{figure-gaugefix-excited}, the statistical errors for the free-form smeared effective mass are smaller.

The bottomonium ground state and excited state energies are extracted by multi-correlator multi-exponential fits of the form
\begin{align}
C^i(t)=\sum_n A_n^i\, e^{-E_n t} \quad ,
\end{align}
where two free-form smeared correlators (one optimized for the ground state and the other for the first-excited state) and a local operator were fit simultaneously to, typically, five exponentials. The analysis was repeated using gauge-fixed wave function smeared correlators. Ratios of the statistical errors for the two methods $\frac{\sigma_\text{gauge-fixed}}{\sigma_\text{free-form}}$ are given in Table~\ref{table-ratio}; statistical uncertainties from free-form smearing are consistently smaller in all channels.

\begin{table}[htb]
\centering
\label{table-ratio}
\caption{Ratios of statistical errors $\frac{\sigma_\text{gauge-fixed}}{\sigma_\text{free-form}}$ for ground state and first-excited state energies extracted by multi-exponential fits. All ratios are greater than one; free-form smearing consistently has smaller errors than gauge-fixed wave function smearing.}
\begin{tabular}{ccc}
& ground state & first-excited state \\
\hline
$^1S_0$ & $1.1$ & $1.4$ \\
$^3S_1$ & $1.2$ & $1.3$ \\
$^1P_1$ & $1.7$ & $2.6$ \\
$^3P_0$ & $1.3$ & $2.1$ \\
$^3P_1$ & $1.4$ & $2.1$ \\
$^3P_2$ $E$ & $1.8$ & $2.0$ \\
$^3P_2$ $T_2$ & $1.6$ & $2.2$ \\
\end{tabular}
\begin{tabular}{ccc}
& ground state & first-excited state \\
\hline
$^1D_2$ $E$ & $1.7$ & $1.4$ \\
$^1D_2$ $T_2$ & $1.7$ & $1.8$ \\
$^3D_2$ $E$ & $1.3$ & $1.2$ \\
$^3D_2$ $T_2$ & $1.7$ & $1.3$ \\
$^3D_3$ $A_2$ & $2.7$ & $2.4$ \\
$^3D_3$ $T_2$ & $2.3$ & $1.7$ \\
&
\end{tabular}
\end{table}

\section{Bottomonium Spectrum from Free-form Smearing}\label{sec:spectrum}

\begin{figure}[htb]
\centering
\vspace{6mm}
\includegraphics[scale=0.5]{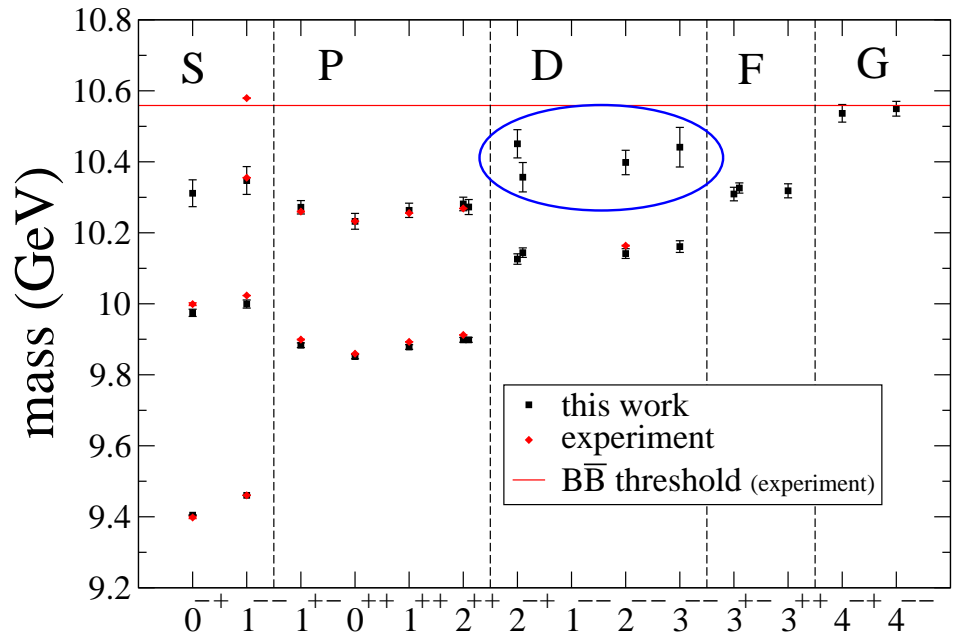}
\caption{Bottomonium spectrum extracted from simultaneous multi-exponential fits of free-form smeared correlation functions and current experimental values \cite{PDG}. Highlighted (inside the blue oval) are first lattice results of bottomonium D-wave first-excited states.}
\label{figure-spectrum}
\end{figure}

The bottomonium spectrum is extracted by simultaneous multi-correlator multi-exponential fits of free-form smeared correlation functions. A local operator, which contains a mixture of the low lying states, is also included in the fit. Reliable fit values are obtained for the S-wave second excited state, S, P and D-wave first excited states and S, P, D, F, and G-wave ground state energies. The NRQCD energies are converted to masses in physical units by fixing the upsilon S-wave ground state to its experimental value \cite{PDG} and adding the necessary energy shift given by $M=M^\text{exp}(\Upsilon(1S))+a^{-1}(aE^\text{sim}-aE^\text{sim}(\Upsilon(1S)))$, where $aE^\text{sim}$ is the dimensionless energy extracted from simulations. The bottomonium spectrum from free-form smeared correlators in shown in Fig.~\ref{figure-spectrum}.

First results from lattice simulations for the D-wave first-excited state are highlighted in Fig.~\ref{figure-spectrum}. They are below the $B\overline{B}$-threshold and consistent with a quark model prediction of 10.45 GeV \cite{Godfrey:1985xj}. Clean first-excited state signals from free-form smeared correlators, as seen in Fig.~\ref{figure-gaugefix-excited}, were necessary to get a robust fit of the D-wave first excitation.

\section{Conclusions}\label{sec:conclusions}

Free-form smearing is an excellent method to extract the spectrum of bottomonium, and has been used to obtain a first lattice result for the mass of the D-wave first-excited state. Free-form smearing gives smaller statistical errors than the gauge-fixed smearing method and a cleaner ground state signal than the gauge-invariant gaussian smearing technique. Further work is required to apply free-form smearing to a full correlator matrix.

\section*{Acknowledgments}

The authors thank Georg von Hippel for helpful discussions about free-form smearing.
This work was supported in part by the Natural Sciences and
Engineering Research Council (NSERC) of Canada, and by computing resources of
WestGrid\cite{westgrid} and SHARCNET\cite{sharcnet}.

\end{document}